\documentclass[a4paper,10pt,twoside]{cpc-hepnp}

\usepackage{multicol}
\usepackage{graphicx}
\usepackage{booktabs}
\usepackage{amssymb,bm,mathrsfs,bbm,amscd}
\usepackage[tbtags]{amsmath}
\usepackage{lastpage}

\def\nuc#1#2{\relax\ifmmode{}^{#1}{\protect\text{#2}}\else${}^{#1}$#2\fi}

\begin{document}

\fancyhead[c]{\small Chinese Physics C~~~Vol. 37, No. 1 (2013) 010201}
\fancyfoot[C]{\small 010201-\thepage}

\footnotetext[0]{Received 27 March 2013}

\title{Effects of relativistic kinematics in heavy ion elastic scattering
\thanks{Supported by National Natural Science
Foundation of China (11275018, 11021504, and 11035001), and by the Research Fund
for the Doctoral Program of Higher Education of China (No. 20121102120026))
}}

\author{%
      PANG Dan-Yang$^{1,2)}$\email{dypang@buaa.edu.cn}%
}
\maketitle

\address{%
$^1$ School of Physics and Nuclear Energy Engineering, Beihang University,
Beijing, 100191, China, and\\
$^2$ International research center for nuclei and particles in the cosmos, Beihang University, Beijing, 100191, China
}

\begin{abstract}
Relativistic corrections to the reaction kinematic parameters were made for
elastic scattering of \nuc{6}{Li}, \nuc{12}{C} and \nuc{40}{Ar} from \nuc{40}{Ca},
\nuc{90}{Zr}, and \nuc{208}{Pb} targets at incident energies between 20 and 100
MeV/nucleon. Results of optical model calculations show that the effects of
such corrections are important in describing the angular distributions of
elastic scattering cross sections for heavy ion scattering at incident energies
as low as around 40 MeV/nucleon. The effects on the total reaction cross
sections on the other hand, were found to be small within the energy range
studied when the optical model potential is fixed.
\end{abstract}

\begin{keyword}
keyword,  heavy ion elastic scattering, optical model, relativistic kinematics
\end{keyword}

\begin{pacs}
24.10.Ht, 25.70.-z, 25.75.-q
\end{pacs}

\footnotetext[0]{\hspace*{-3mm}\raisebox{0.3ex}{$\scriptstyle\copyright$}2013
Chinese Physical Society and the Institute of High Energy Physics
of the Chinese Academy of Sciences and the Institute
of Modern Physics of the Chinese Academy of Sciences and IOP Publishing Ltd}%

\begin{multicols}{2}

\section{Introduction}

It is well-known that at sufficiently high incident energies relativistic
effects have to be taken into account in describing heavy ion elastic scattering
\cite{Ingemarsson-PS-1974,Farid-PLB-1984}. There are several aspects in treating the
relativity, for example, the use of the Schr\"{o}dinger equation is not
relativistically correct and, at least, some re-interpolation of the nuclear
optical potential should be made \cite{Nadesen-PRC-1981}, the parameters of
reaction kinematics, namely, the atomic masses and incident energies have to be
modified. The later aspect is rather simple, but it has been found to be
important for some cases. For example, Farid and Satchler studied the
latter aspect and their effects on
the optical model potentials extracted by fitting experimental data. Changes in
potential strength by larger than 20\% were found for heavy ion elastic
scattering at incident energy as low as 44 MeV/nucleon, at which the relativistic
effects were usually thought to be
negligible \cite{Farid-PLB-1984}. The effects on shapes of angular distributions of elastic
scattering cross sections, however, have not been reported yet.

Recently, in our study of systematic heavy ion potentials we also found the
necessity of introducing the relativistic corrections to the reaction
kinematic parameters when the incident energy of a projectile nucleus is above 40 MeV/nucleon
\cite{Xu-PRC-2013}. One example is shown in Fig.\ref{fig-17O-40Ar-rel-effect},
where optical model calculations with and without relativistic corrections were
compared with the experimental data. Clearly one can see that the inclusion of
relativistic corrections helps to improve the description of experimental data.
In this paper, we examine the effect of relativistic corrections to reaction
kinematic parameters on heavy ion elastic scattering and total reaction cross
sections, and study their dependence on the projectile and target masses and
incident energies. For such purposes, \nuc{6}{Li}, \nuc{12}{C} and
\nuc{40}{Ar} elastic scattering from \nuc{40}{Ca}, \nuc{90}{Zr} and
\nuc{208}{Pb} are studied at incident energies between 10 and 100 MeV/nucleon,
which is the range where the relativistic effects were usually thought to be
negligible.

\begin{center}
\includegraphics[width=0.4\textwidth]{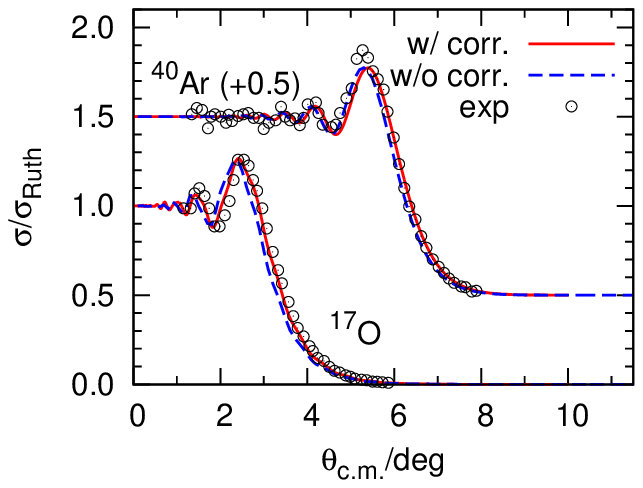}
\figcaption{\label{fig-17O-40Ar-rel-effect} (Color online) Optical model calculations of
\nuc{17}{O} and \nuc{40}{Ar} elastic scattering from \nuc{208}{Pb} at 84 and 44
MeV/nucleon, respectively, with and without taking into account the relativistic
corrections to the reaction kinematic parameters. The experimental data are
from Ref.\cite{Neto-NPA-1993, Alamanos-PLB-1984}.}
\end{center}

This paper is organized as the following: the relativistic corrections to the
parameters of reaction kinematics are introduced in Section {\ref{sect-formula}}.
Results of optical model calculations with and without taking into account such
corrections are shown in Section {\ref{sect-calculations}}, and the summary of
this paper is given in Section {\ref{sect-summary}}.

\section{Effects of relativistic corrections to reaction kinematic parameters}
\label{sect-formula}

In Ref.\cite{Satchler-NPA-1992}, Satchler demonstrates that the
Klein-Gorden equation describing the elastic scattering of a pion from an
atomic nucleus can be transformed into a Schr\"{o}dinger equation by modifying
the mass number (in atomic mass unit) and incident energy of the pion
projectile. Since the mass of pion is much smaller than that of the target
nucleus, correction to target mass is neglected. Such prescription can be
generalized to cases of nucleus-nucleus scattering by taking into account the
projectile-target symmetry. For such cases, the masses of the projectile and the
target nuclei, $m_\textrm{p}$ and $m_\textrm{t}$, respectively, and the effective bombarding
energy, $E_\text{L}$ should be translated as\cite{Satchler-NPA-1992}:
\begin{eqnarray}
M_\textrm{p} &=& m_\textrm{p}\gamma_\textrm{p},\\
M_\textrm{t} &=& m_\textrm{t}\gamma_\textrm{t},\\
E_\textrm{L} &=& E_\textrm{c.m.}M_\textrm{p}/\mu,
\end{eqnarray}
where $\mu$ is the reduced mass
\begin{equation}
 \mu=\frac{M_\textrm{p}M_\textrm{t}}{M_\textrm{p}+M_\textrm{t}}.
\end{equation}
The relativistically correct center-of-mass energy $E_\textrm{c.m.}$ is
\begin{equation}
 E_\textrm{c.m.}= \frac{(k\hbar c)^2}{2\mu},
\end{equation}
in which, $c$ is the speed of light, $k$ is the wave number:
\begin{equation}
 k= \frac{m_\textrm{p}  c^2}{\hbar c}(\gamma_\textrm{p}^2-1)^{1/2}
\end{equation}
and
\begin{eqnarray}
 \gamma_\textrm{p} &=& \frac{x_\textrm{p}+\gamma_L}{\sqrt{1+x_\textrm{p}^2+2x_\textrm{p}\gamma_L}},\\
 \gamma_\textrm{t} &=& \frac{x_\textrm{t}+\gamma_L}{\sqrt{1+x_\textrm{t}^2+2x_\textrm{t}\gamma_L}},
\end{eqnarray}
with
\begin{equation}
x_\textrm{p} = \frac{m_\textrm{p}}{m_\textrm{t}}, \, x_\textrm{t}=\frac{m_\textrm{t}}{m_\textrm{p}}, \, \textrm{and }
\gamma_L=1+\frac{E_\textrm{lab}}{m_\textrm{p} c^2}.
\end{equation}
This prescription has been extensively used in, e.g.,
Ref.\cite{Youngblood-PRC-1997, Satchler-PRC-1997,Clark-PRC-1998,
Brandan-NPA-2001}.

\end{multicols}

\begin{center}
\includegraphics[width=16cm]{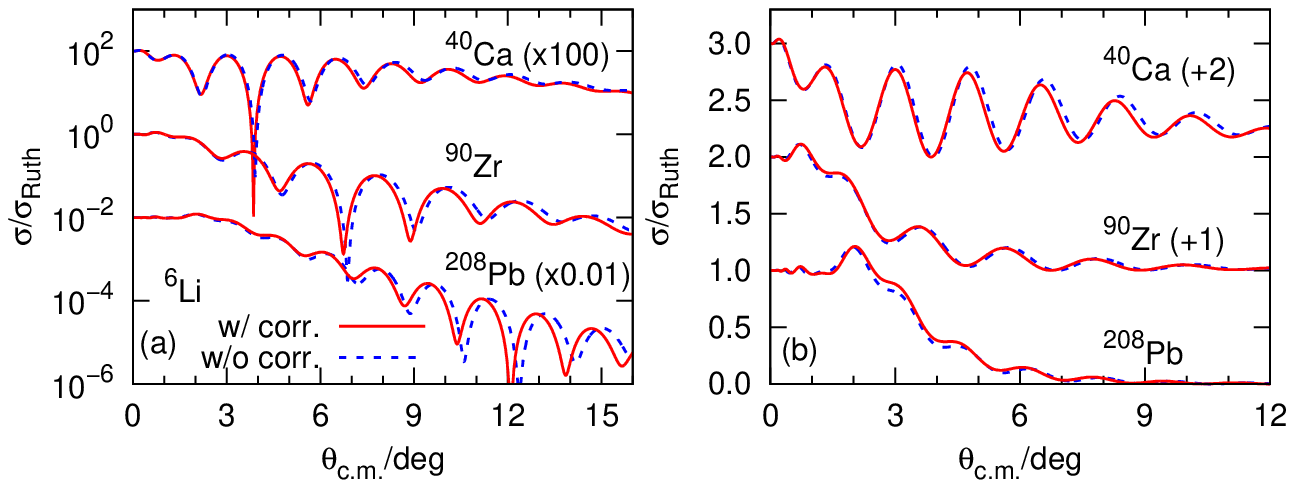}
\figcaption{\label{fig-rel-effect-el-6Li} (Color online) Angular distributions of \nuc{6}{Li}
elastic scattering from \nuc{40}{Ca}, \nuc{90}{Zr} and \nuc{208}{Pb} at 80
MeV/nucleon shown in (a) logarithmic scale and (b) linear scale, from optical
model calculations with and without taking into account the relativistic
corrections to the reaction kinematic parameters.}
\end{center}

\begin{multicols}{2}

\section{Results of optical model calculations}\label{sect-calculations}

Recently a systematic nucleus-nucleus potential was proposed with a
single-folding model approach, with which a heavy ion potential is calculated by
folding the nucleon-nucleus potential with the nucleon density distributions of
the projectile \cite{Pang-PRC-2011, Pang-JPG-2012}. The Bruy\`{e}res
Jeukenne-Lejeune-Mahaux (JLMB) model nucleon-nucleus potential
\cite{Bauge-PRC-1998,Bauge-PRC-2001} was used and the systematics of potential
parameters was found for heavy targets ($A\gtrsim40$) with incident energies
from 5 to 40 MeV/nucleon. This potential was found to give reasonable account of
heavy ion potential even up to 100 MeV/nucleon \cite{Xu-PRC-2013}. With this
systematic potential we are now able to study the effects of relativistic
corrections to the reaction kinematic parameters and their dependence on
projectile and target masses and incident energies realistically. Figures
\ref{fig-rel-effect-el-6Li} and \ref{fig-rel-effect-el-40Ar} show the angular
distributions of elastic scattering cross sections as ratio to the Rutherford
cross sections for \nuc{6}{Li} and \nuc{40}{Ar} projectiles, respectively, from
\nuc{40}{Ca}, \nuc{90}{Zr} and \nuc{208}{Pb} at 80 MeV/nucleon, with and without
taking into account the relativistic corrections to the reaction kinematic
parameters. Clearly, one can see that, for a fixed incident energy, the importance
of the above relativistic corrections increases with the increase of the mass of
the target nucleus.
\end{multicols}

\begin{center}
\includegraphics[width=16cm]{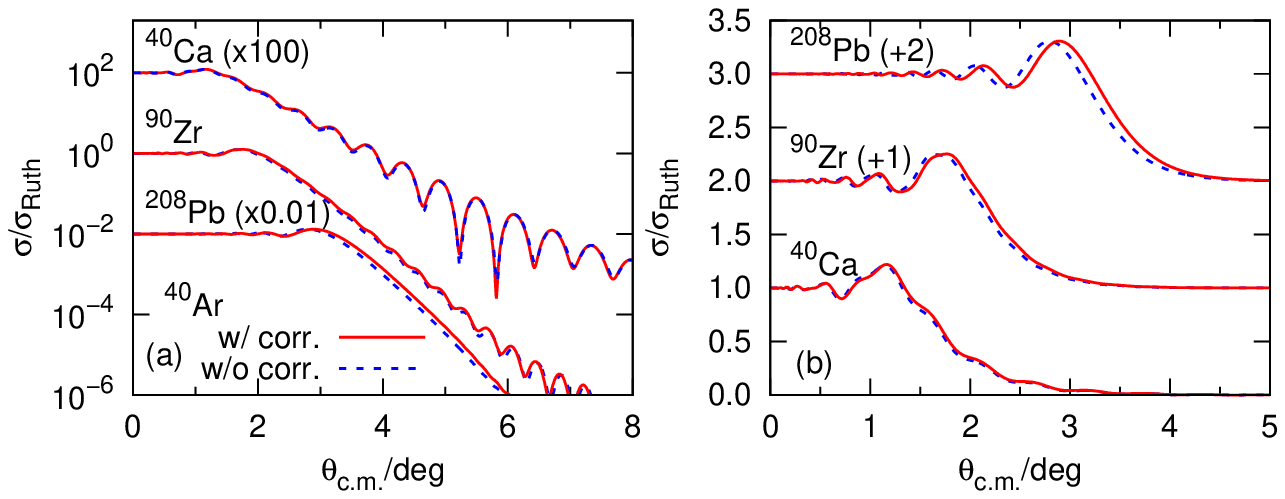}
\figcaption{\label{fig-rel-effect-el-40Ar} (Color online) The same as
Fig.\ref{fig-rel-effect-el-6Li} but for the \nuc{40}{Ar} projectile.}
\end{center}

\begin{center}
\includegraphics[width=16cm]{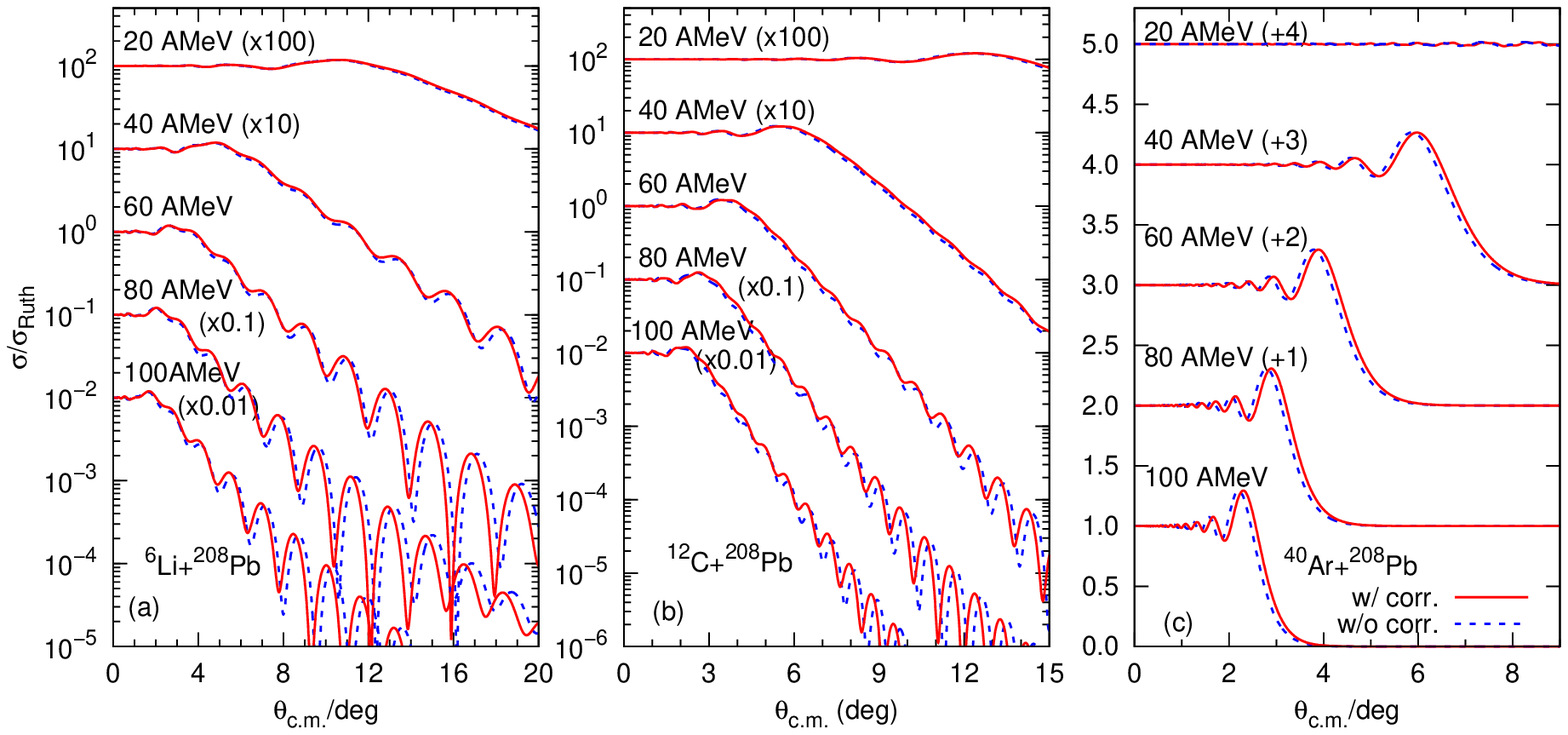}
\figcaption{\label{fig-rel-effect-el-energy} (Color online) Angular distributions of (a)
\nuc{6}{Li}, (b) \nuc{12}{C}, and (c) \nuc{40}{Ar} elastic scattering from
\nuc{208}{Pb} at 20, 40, 60, 80 and 100 MeV/nucleon from optical model
calculations with and without taking into account the relativistic corrections
to the reaction kinematic parameters.}
\end{center}

\begin{multicols}{2}
In order to study how the importance of the relativistic corrections evolves
with the incident energy of the projectile we study \nuc{6}{Li}, \nuc{12}{C}
and \nuc{40}{Ar} elastic scattering from \nuc{208}{Pb} at 20, 40, 60, 80 and 100
MeV/nucleon. The results are shown in Fig.\ref{fig-rel-effect-el-energy}. Two
types of angular distributions are clearly shown: (i) the Fraunhofer-like
scattering, which shows oscillations at large angles when the Coulomb potential
is not so strong compared with the nuclear potential (i.e., $\eta\lesssim 1$,
where $\eta=Z_\textrm{p}Z_\textrm{t}e^2\mu/\hbar^2k$ is the
Sommerfeld parameter with $Z_\textrm{p}$ and $Z_\textrm{t}$ being the charge numbers
of the projectile and target nuclei, and $\mu$ being the reduced mass), and (ii) the  Fresnel-like scattering, which does not
show oscillations at large angles when the Coulomb potential is strong ($\eta\gg1$)
\cite{Satchler-book}.

It is well known that the separation of successive maxima or minima of the Fraunhofer-like scattering, $\Delta\theta$, relates with the
grazing angular momentum $\lambda_g$ and the wave number $k$ by:
\begin{equation}
\Delta\theta \simeq \frac{\pi}{\lambda_g} \simeq \frac{\pi}{kR_g},
\end{equation}
where $R_g$ is the critical or grazing radius at which the projectile and target
nuclei begin to experience the strong nuclear interaction acting between
them\cite{Satchler-book}. We can thus quantify the effect of relativistic
corrections to Fraunhofer-like scattering by the change of wave numbers $k$. The
result is shown in Fig.\ref{fig-rel-effect-k}, where the ratio $k/k^{rel}$ is
plotted as a function of the incident energy for \nuc{6}{Li}, \nuc{12}{C} and
\nuc{40}{Ar} projectiles with \nuc{40}{Ca}, \nuc{90}{Zr} and \nuc{208}{Pb}
targets with $k^{rel}$ and $k$ being the wave numbers calculated with and
without taking into account the relativistic corrections. One can see that the
change in $k$ decreases with the increase of projectile mass and increases with
the increase of the target mass, which is consistent with the observation in angular
distributions shown in Figs. \ref{fig-rel-effect-el-6Li},
\ref{fig-rel-effect-el-40Ar} and \ref{fig-rel-effect-el-energy}.

\begin{center}
\includegraphics[width=0.4\textwidth]{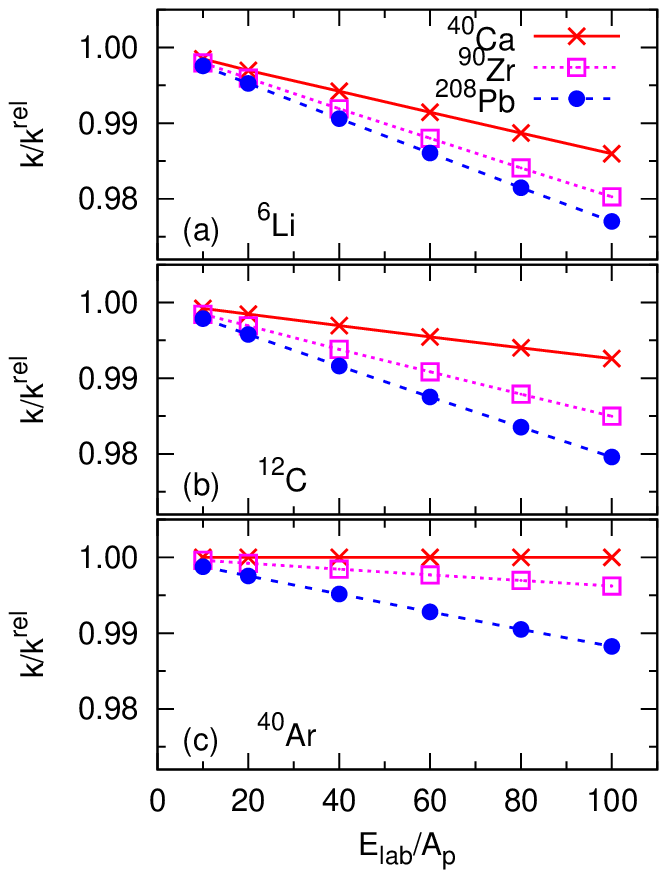}
\figcaption{\label{fig-rel-effect-k} (Color online) Effect of the relativistic corrections on
wave number $k$ as a function of incident energies with target nuclei
\nuc{40}{Ca}, \nuc{90}{Zr} and \nuc{208}{Pb} for (a) \nuc{6}{Li} and (b)
\nuc{12}{C} and (c) \nuc{40}{Ar}.}
\end{center}

A Fresnel-like
scattering, on the other hand, does not show many oscillations in the angular
distributions. It is characterized with the quarter-point angle
$\theta_{\frac{1}{4}}$, which relates the wave number $k$ by
\cite{Satchler-book}:
\begin{equation}
kR_{\frac{1}{4}}=\eta\left[1+\textrm{cosec}\left(\frac{1}{2}\theta_{\frac{1}{4}}
\right)\right],
\end{equation}
where $R_{\frac{1}{4}}$ is the interaction radius which is taken to be the
closest approach for the orbit $L_{\frac{1}{4}}$ which satisfies
\begin{equation}
L_{\frac{1}{4}}+\frac{1}{2}=\eta\cot\left(\frac{1}{2}\theta_{\frac{1}{4}}
\right).
\end{equation}
For such Fresnel-like scattering, we choose to quantify the effect of
relativistic corrections by changes in $\theta_{\frac{1}{4}}$. The results of
optical model calculations show that for \nuc{40}{Ar} elastic scattering from
\nuc{208}{Pb} at 20, 40, 60, 80 and 100 MeV/nucleon, the changes in
$\theta_{\frac{1}{4}}$ are 1\%, 1.5\%, 2.2\%, 2.9\% and 3.6\%, respectively,
which can be seen in Fig.\ref{fig-rel-effect-el-energy}(c). Note that such
amount of changes are distinguishable with experimental data, as can be seen
from Fig.\ref{fig-17O-40Ar-rel-effect}.

\begin{center}
\includegraphics[width=0.4\textwidth]{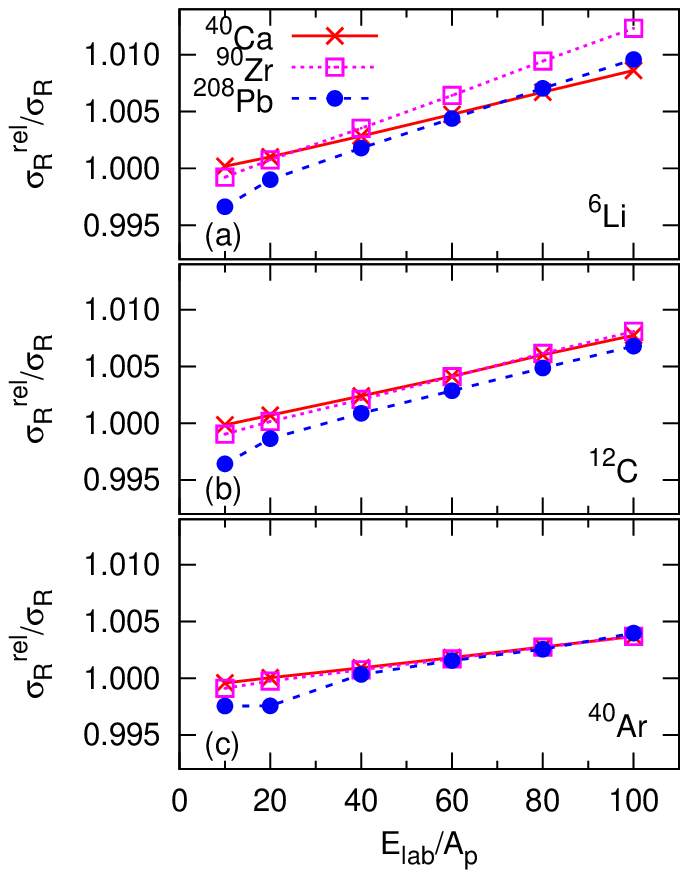}
\figcaption{\label{fig-rel-effect-xsec} (Color online) Effect of relativistic corrections on
total reaction cross sections from optical model calculations as a function of
incident energies with target nuclei \nuc{40}{Ca}, \nuc{90}{Zr} and
\nuc{208}{Pb} for (a) \nuc{6}{Li} and (b) \nuc{12}{C}, and (c) \nuc{40}{Ar}.}
\end{center}

We also evaluate the effect of the above relativistic corrections on the total
reaction cross sections $\sigma_R$ for projectiles \nuc{6}{Li}, \nuc{12}{C} and
\nuc{40}{Ar} with \nuc{40}{Ca}, \nuc{90}{Zr} and \nuc{208}{Pb} targets. The
results are shown in Fig.\ref{fig-rel-effect-xsec} as ratios
$\sigma_R^\textrm{rel}/\sigma_R$, where $\sigma_R^{rel}$ and $\sigma_R$ are the
total reaction cross sections obtained with and without taking into account the
relativistic corrections in the optical model calculations. For most of the cases
we studied the changes in $\sigma_R$ are within 0.5\%, which is negligible
compared with the typical experimental uncertainties in a total reaction cross
section measurement. Note that the authors of Ref.\cite{Farid-PLB-1984}
found that the effect of relativistic corrections to the total reaction cross
section of \nuc{40}{Ar} and \nuc{60}{Ni}, \nuc{120}{Sn} and \nuc{208}{Pb}
targets at 44 MeV/nucleon were 0.2\%, 1.2\% and 2.2\%, respectively, which are
much larger than the results shown in Fig.\ref{fig-rel-effect-xsec}. There is no
conflict between the results in the present paper and those in
Ref.\cite{Farid-PLB-1984} because in this work the optical model potentials were
fixed. This is the reason why the changes in the total reaction cross sections are
small. In Ref.\cite{Farid-PLB-1984} the reported changes in $\sigma_R$
were due to the use of two different sets of optical model potentials obtained by
fitting the experimental data with and without applying the relativistic
corrections to the reaction kinematics.

\section{Summary}\label{sect-summary}

In summary, \nuc{6}{Li}, \nuc{12}{C} and \nuc{40}{Ar} elastic scattering from
\nuc{40}{Ca}, \nuc{90}{Zr} and \nuc{208}{Pb} with incident energies between 20
and 100 MeV/nucleon were studied with optical model calculations. The effects of
relativistic corrections to the reaction kinematic parameters were studied for
their angular distributions and the total reaction cross sections. The results of
the calculations show that the relativistic corrections are important for
describing the angular distributions of heavy ion elastic scattering cross
sections at incident energies as low as around 40 MeV/nucleon. The total
reaction cross sections are found to be little affected by such corrections.
Besides the scattering problems studied in this paper,
it might be interesting to study how important these relativistic effects are in
decay problems, such as $\alpha$ decay studied in Ref.\cite{XuChang-PRC-2006}.

\end{multicols}

\vspace{-1mm}
\centerline{\rule{80mm}{0.1pt}}
\vspace{2mm}

\begin{multicols}{2}

\end{multicols}

\clearpage

\end{document}